\documentclass[%
 reprint,
 amsmath,amssymb,
 aps,
]{revtex4-2}

\usepackage{graphicx}
\usepackage{dcolumn}
\usepackage{bm}
\usepackage{xcolor} 

\begin{document}

\preprint{APS/123-QED}

\title{Influence of transverse oscillatory shear on the viscous fingering instability}



\author{Zhaoning Liu}
\affiliation{%
 Department of Physics and The James Franck and Enrico Fermi Institutes, University of Chicago, Chicago IL, 60637
}%

\date{\today}

\begin{abstract}
Oscillatory shear can delay the onset of viscous fingering. In addition, uniaxial translational shear also breaks the symmetry of the resulting pattern so that the interface evolves differently depending on its direction of propagation relative to the shear axis. We observe a delayed fingering onset and reduced subsequent finger growth rate for all directions as shear speed or amplitude increases. However, shear widens fingers growing in the direction perpendicular to shear and elongates those in the parallel directions. After onset, neighboring fingers in the perpendicular directions can reconnect near their base when the shear amplitude exceeds approximately five times the excess linear finger wavelength at the onset. Concentration profiles reveal smoother azimuthal edges and overlapping structures behind the tips. These observations show that shear smears individual fingers and connects neighboring ones. This correlates lateral fingering interactions with radial propagation.

\end{abstract}


\maketitle
\onecolumngrid

\begin{figure}[h]
    \centering
    \includegraphics[width=0.9\linewidth]{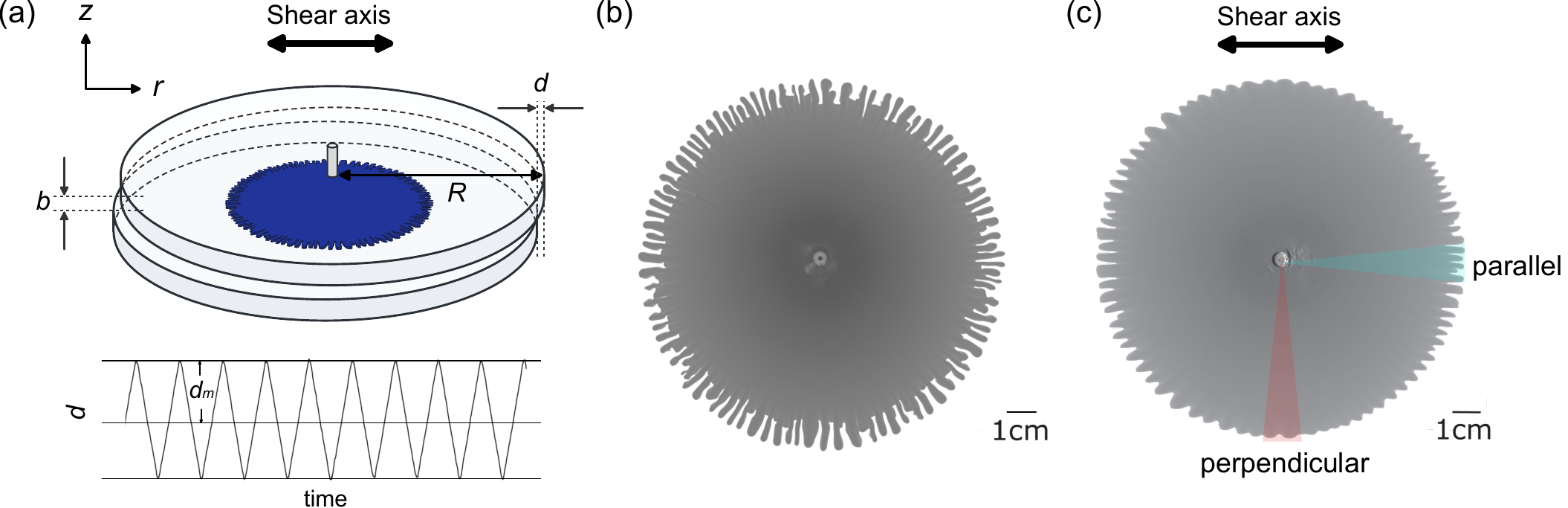}
    \caption{Viscous fingering with translational oscillatory shear. (a) Schematic of a radial Hele-Shaw cell with plates of radius $R$ separated by a gap $b$ with $b\ll R$. The coordinate $z$ spans the gap: $0\leq z \leq b$. The top plate can be sheared uniaxially with respect to the fixed bottom plate with a displacement $d$. The displacement $d$ versus time shown in the bottom plot has a triangular waveform with amplitude $d_\mathrm{m}$ and velocity $\pm V_\mathrm{s}$. For experiments without shear, $d=0$. The less viscous inner fluid (blue) is injected from a central inlet in the top plate. Top view of finger formation (b) without and (c) with shear using $d_\mathrm{m}=\mathrm{6~mm}$, $V_\mathrm{s}=\mathrm{3.1~mm/s}$. In (c), shaded wedges mark fingers propagating in the direction parallel (blue) and perpendicular (red) to the shear axis. Fingers in all directions are suppressed compared to the $d=0$ case in (b). In (b,c), $b=\mathrm{305~\mu m}$, $Q=\mathrm{133~\mu L/s}$, $\eta_\mathrm{in}=\mathrm{35\pm 2~mPa\cdot s}$, $\eta_\mathrm{out}=\mathrm{218\pm 5~mPa\cdot s}$.}
    \label{fig:setup}
\end{figure}

\twocolumngrid
\section{Introduction}

Many pattern-forming systems develop structures that depend on the nature of the imposed driving~\cite{cross1993}. Introducing an additional, independently controlled, external drive provides a way to investigate not only the physical mechanisms underlying the pattern formation but also the response of patterns to an external perturbation. 

A prototypical example of pattern formation where an additional perturbation can be applied is the viscous fingering instability in miscible fluids. When a less viscous fluid displaces a more viscous one in a confined geometry, \textit{e.g.}, the narrow gap between two parallel plates in a Hele-Shaw cell  (see Fig.~\ref{fig:setup}a), an initially smooth interface can develop protrusions that develop into fingers~\cite{saffman1958,Homsy1987} as shown in Fig.~\ref{fig:setup}b. In this case, these fingers produce complex patterns through their growth and interaction~\cite{paterson1981,Chui2015.92.041003}. The directly observable interface makes this system well suited for investigating how an external perturbation changes the patterns that form~\cite{Langer1989,Qing2023}.


Various external perturbations have been explored to alter viscous fingering patterns. Time-dependent injection rates~\cite{Li2009,Dias2012} and time-dependent gap widths~\cite{zheng2015,Pihler2012} can suppress fingering or control the number of fingers that form, and electrokinetic forcing~\cite{MirzadehBazant2017,gao2019} provides further means to modify the instability. Recently, translational oscillatory shear has been used to suppress the instability that occurs in radial displacement with miscible pairs of fluids~\cite{liu2026}. This was correlated with a smoother gap-averaged concentration profile in front of the finger in the direction between the confining plates, $z$ (see Fig.~\ref{fig:setup}a). Shear therefore serves both as a new way to control the pattern formation and as a probe of the mechanism underlying the instability.

Uniaxial shear, however, breaks the rotational symmetry of radial displacement. For fingers propagating parallel to the shear axis, the imposed motion acts mainly along the direction of growth. In contrast, for fingers propagating perpendicular to the axis, it acts mainly in the transverse direction (\textit{i.e.}, across the fingers) so that it deforms the invading fluid laterally. This lateral effect is important because the development of a fingering pattern depends on interactions between neighboring fingers. Nonlinear studies have shown that lateral spreading, shielding, and competition influence which fingers continue to advance~\cite{TanHomsy1988,MirandaWidom1998}, and collective instabilities of finger arrays have been linked to coalescence~\cite{KesslerLevine1986}. Radial growth therefore cannot be understood in terms of the isolated radial direction alone, and the lateral interaction of neighboring fingers also matters. Transverse shear offers a way to study these interactions directly. In particular, it raises the question of how shear-induced lateral deformation modifies the interactions between neighboring fingers, and how these changes in turn affect the radial development of the pattern.

In this work, we investigate these questions by studying fingers propagating perpendicular to the shear axis. We find that oscillatory shear not only delays the fingering onset and reduces subsequent finger growth, but also  widens them laterally as can be seen in Fig.~\ref{fig:setup}c. Measurements of the concentration of the inner fluid show smeared azimuthal profiles with more gradual edges, as well as additional overlapping structure behind the tips. Under a subset of shear parameters, neighboring fingers reconnect, reducing the finger length and its subsequent growth. These observations connect the perpendicular perturbation with changes in lateral morphology and neighboring interactions.  It thus provides a physical example of how lateral deformation influences pattern development in the radial direction.

\section{\label{sec:level1}Results}

\subsection*{Experimental platform}

We use the radial Hele-Shaw apparatus and experimental procedures described in Ref.~\cite{liu2026}. Two glass plates of radius $R=14~\mathrm{cm}$ confine the fluids within a uniform horizontal gap $b=305~\mathrm{\mu m}$ (Fig.~\ref{fig:setup}a). The bottom plate remains stationary while the top plate oscillates along a fixed axis while keeping the gap constant. Its measured displacement $d(t)$ follows a triangular waveform with a maximum fixed amplitude $d_\mathrm{m}$, with a constant speed $\pm V_\mathrm{s}$ (Fig.~\ref{fig:setup}a, bottom). In different experiments, we varied shear amplitude from $d_\mathrm{m} = 0.3~\mathrm{mm}$ to $d_\mathrm{m} = 6~\mathrm{mm}$, and the shear speed from $V_\mathrm{s} = 1.4~\mathrm{mm/s}$ to $V_\mathrm{s} = 13.7~\mathrm{mm/s}$.

The gap is initially filled with the more viscous glycerol–water mixture ($\eta_{\rm out} = 218 \pm 5$ mPa·s). A less viscous glycerol–water mixture ($\eta_{\rm in} = 35 \pm 2$ mPa·s) is then injected into the more viscous one at a constant volumetric rate $Q$, which ranges from $Q = 67~\mathrm{\mu L/s}$ to $Q = 267~\mathrm{\mu L/s}$ across experiments.

Because the inner fluid is dyed, its concentration profile, $C$, can be measured from the transmitted light intensity as discussed in \cite{liu2026,videbaek2019}. As the interface propagates radially outward, the speed decreases as $U=Q/(2\pi br)$. There is diffusion between the two miscible fluids that acts to blur the profile. The effective interfluid diffusivity, $D$, is estimated using Fickian diffusion of glycerol-water mixtures to be $1.21\times 10^{-10}~\mathrm{m^2/s}$.  We ensure that there is a well-defined fluid interface by only tracking the interface before $\mathrm{Pe}\equiv Ub/D$ drops below $1750$, that is, $U\geq 0.7~\mathrm{mm/s}$. 

We define the shear axis as being along $\theta=0$ and $\pi$ directions, and examine perpendicular fingers near $\theta=\pi/2$ and $3\pi/2$. These are shown as the red sectors in Fig.~\ref{fig:setup}c. Parallel fingers propagating along $\theta=0$ and $\pi$ directions are shown as the blue sectors.  Further details of the experiment and analysis methods are given in the Methods section in the Supplementary Information.

This paper focuses on fingers growing in the direction perpendicular to the shear axis. To investigate how the application of shear suppresses the instability, we measured the onset radius, $R_\mathrm{on}$, of the fingering instability, and the subsequent early time finger growth rate, $\Gamma\equiv dR_\mathrm{f}/dR_\mathrm{tip}$, where $R_\mathrm{tip}$ is the radial coordinate for the finger tip and $R_\mathrm{f}$ is the finger length (the radial distance between $R_\mathrm{tip}$ and the valley behind it).  We track finger growth rate before $R_\mathrm{tip}-R_\mathrm{on}$ reaches $10~\mathrm{mm}$ or before the fingers reconnect, whichever comes earlier. As shown in Fig.~\ref{fig:SI_growth}, $R_\mathrm{f}$ increases linearly with $R_\mathrm{tip}$ in the region we track. We measured finger width $W$ and the concentration profile, $C$, of the inner fluid to investigate how the fingers are smeared and reconnected. We propose that the reconnection can be one route by which shear suppresses the instability in the perpendicular direction.
 
\subsection*{Delayed onset and reduced finger growth rate}

\begin{figure}
    \centering
    \includegraphics[width=1.02\linewidth]{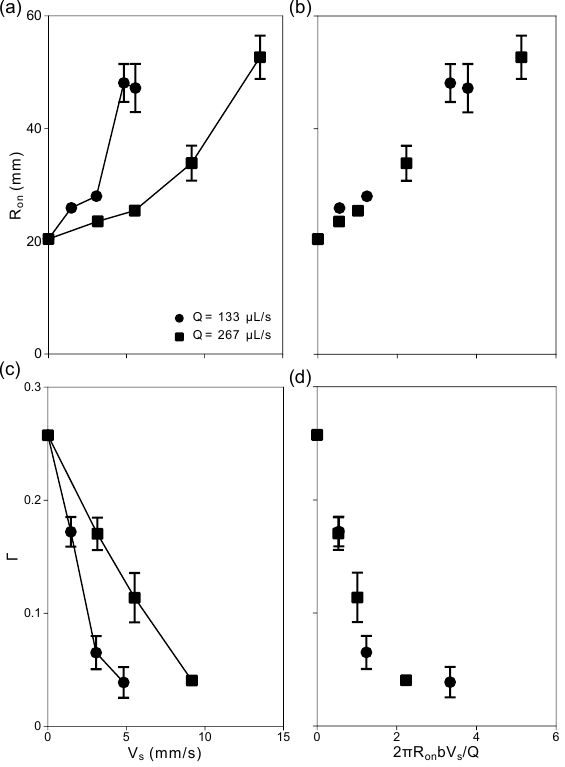}
    \caption{Effect of shear speed on the instability onset and growth rate for fingers perpendicular to the shear axis at a fixed shear amplitude. (a) Onset radius, $R_\mathrm{on}$, and (c) growth rate, $\Gamma$, versus shear speed, $V_\mathrm{s}$. Higher shear speed delays the onset and reduces growth rate; the effect is stronger at the lower injection rate. (b,d) Same data as in (a,c), but $V_\mathrm{s}$ is normalized to be dimensionless by the interface propagation speed at the onset radius. The curves collapse for different $Q$. The shear amplitude is fixed at $d_\mathrm{m}=\mathrm{3~mm}$.}
    \label{fig:effect_vs}
\end{figure}

\begin{figure}
    \centering
    \includegraphics[width=1.02\linewidth]{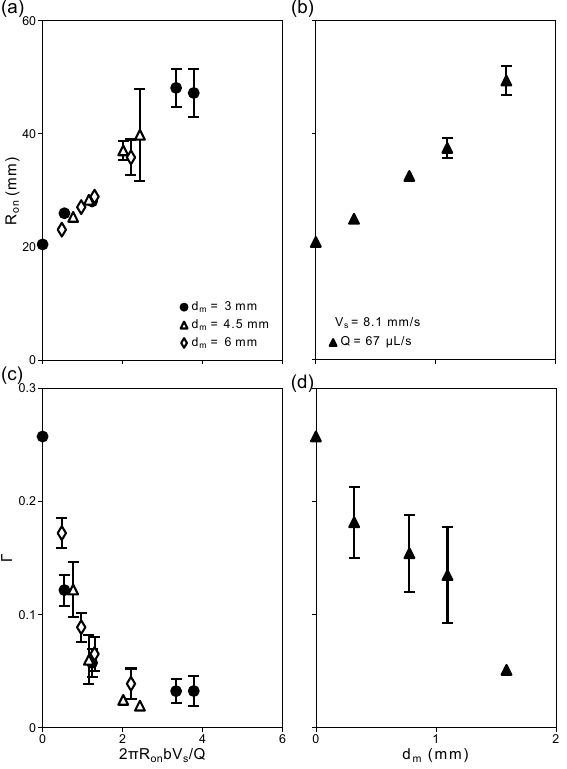}
    \caption{Effect of shear amplitude on the instability onset and growth rate for fingers perpendicular to the shear axis. (a) $R_\mathrm{on}$ and (c)~$\Gamma$ versus the dimensionless shear speed, $2\pi R_\mathrm{on} b V_s/Q$, for different shear amplitudes. Data for all three amplitudes fall on the same curve, so over this range the normalized shear speed alone sets the response. (b) $R_\mathrm{on}$ and (d) $\Gamma$ versus $d_m$ at fixed $V_s = 8.1~\mathrm{mm/s}$ and $Q = 67~\mathrm{\mu L/s}$. At this high $V_\mathrm{s}/Q$, increasing $d_\mathrm{m}$ also more effectively suppresses the instability.
   }
    \label{fig:effect_dm}
\end{figure}

Although in the direction perpendicular to the shear axis, shear does not radially smoothen the interface spanning the thin gap significantly (Fig.~\ref{fig:SI_tip}), the instability (seen both in $R_\mathrm{on}$ and $\Gamma$) is similarly suppressed as shown in Fig.~\ref{fig:effect_vs} and  Fig.~\ref{fig:effect_dm}. As shown in Fig.~\ref{fig:effect_vs}a, $R_\mathrm{on}$ increases with $V_\mathrm{s}$ at fixed $d_\mathrm{m}=3~\mathrm{mm}$. The delay is greater for lower injection rate, $Q$. Correspondingly, Fig.~\ref{fig:effect_vs}c shows that the early growth rate decreases with increasing $V_\mathrm{s}$. When $V_\mathrm{s}$ is scaled to be dimensionless by the interface propagation speed at the onset radius, $2\pi R_\mathrm{on}b V_\mathrm{s}/Q$, the data taken at different $Q$ approximately collapse, as shown in Fig.~\ref{fig:effect_vs}b,d. This shows that the response is due to a competition between imposed shear and radial flow propagation rather than by shear speed alone. As one would expect that there is more diffusion when $Q$ is small, this collapse of data also indicates diffusion is not important in the parameter range studied here.

Figure~\ref{fig:effect_dm} examines the role of shear amplitude. For the speed sweeps at $d_\mathrm{m}=3$,$4.5$, and $6~\mathrm{mm}$, $R_\mathrm{on}$ and $\Gamma$ follow approximately common trends (Fig.~\ref{fig:effect_dm}a,c). In this range, the amplitude dependence is weak compared to the dependence on the shear speed. In contrast, at a larger shear-to-injection ratio $V_\mathrm{s}/Q=8.1/67\simeq0.12~\mathrm{mm}^{-2}$, increasing $d_\mathrm{m}$ also delays the onset and suppresses finger growth (Fig.~\ref{fig:effect_dm}b,d).

These results demonstrate that increasing either shear amplitude or speed also suppresses the instability for fingers perpendicular to the shear axis.

\subsection*{Symmetry breaking and finger widening}

Although the onset of fingering, $R_\mathrm{on}$, and the subsequent finger growth rate, $\Gamma,$ behave similarly in the parallel and perpendicular directions, shear alters finger morphology differently in the two cases. Figure~\ref{fig:setup}c shows an anisotropic pattern where fingers in the perpendicular directions are laterally widened, while those in the parallel directions appear more elongated. 

\begin{figure}
    \centering
    \includegraphics[width=1.02\linewidth]{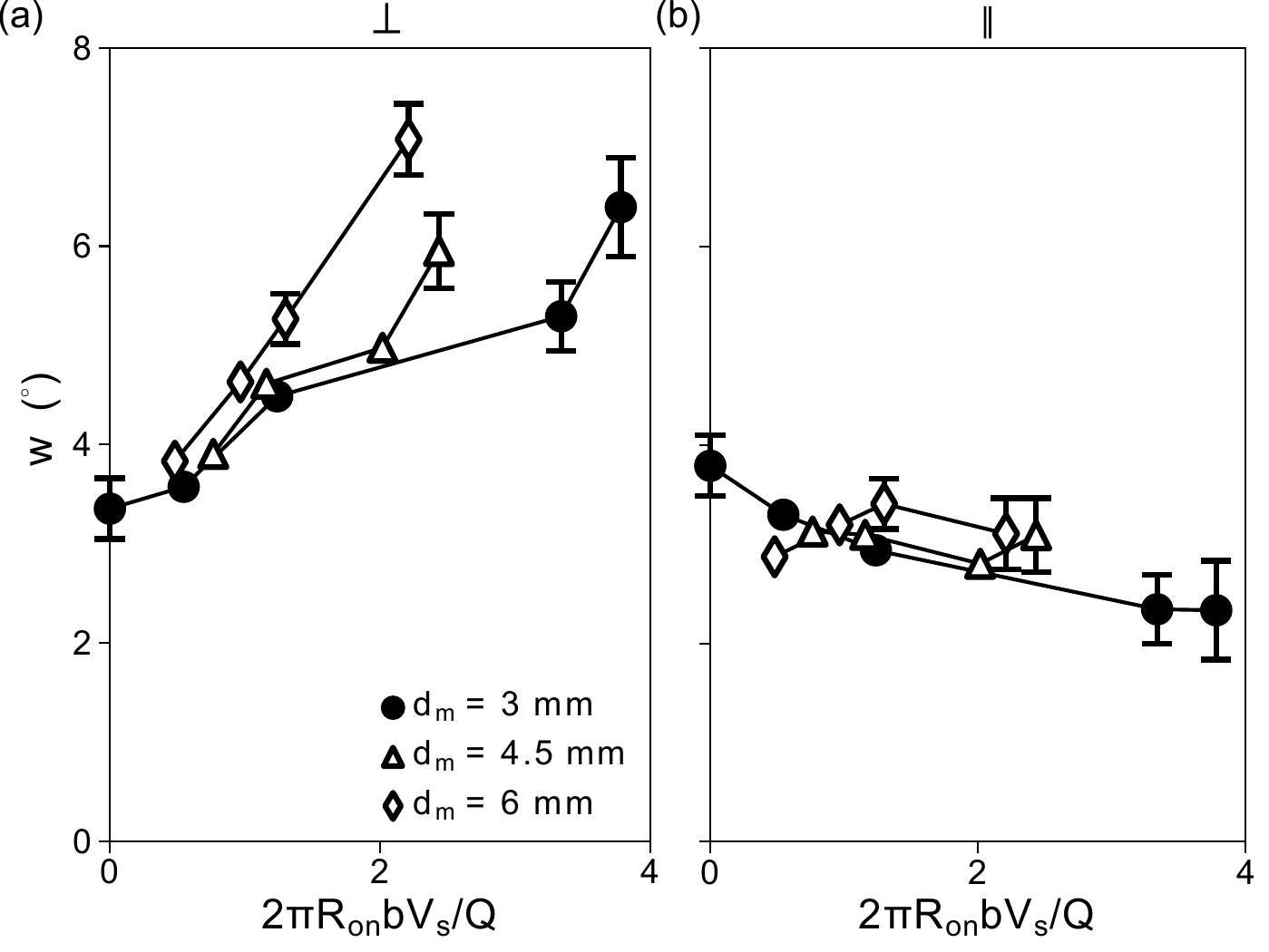}
    \caption{ Fingers grow wider in the perpendicular growth direction but become narrower in the parallel direction. (a) In the perpendicular direction, angular finger width, $W$, increases with dimensionless shear speed. The effect is stronger at higher $d_\mathrm{m}$. (b) Parallel to the shear axis, $W$ decreases as shear speed increases, and the data for all three amplitudes nearly overlap.}
    \label{fig:width}
\end{figure}

This difference is quantified using the angular finger width, $W$. Figure ~\ref{fig:width} shows opposite trends in the two directions. The width in the perpendicular direction increases with shear speed, and this effect is more pronounced at larger $d_\mathrm{m}$ (Fig.~\ref{fig:width}a). In contrast, fingers in the parallel direction narrow with increasing shear speed, with only small variation as the shear amplitude is varied (Fig.~\ref{fig:width}b).

\begin{figure}
    \centering
    \includegraphics[width=1.02\linewidth]{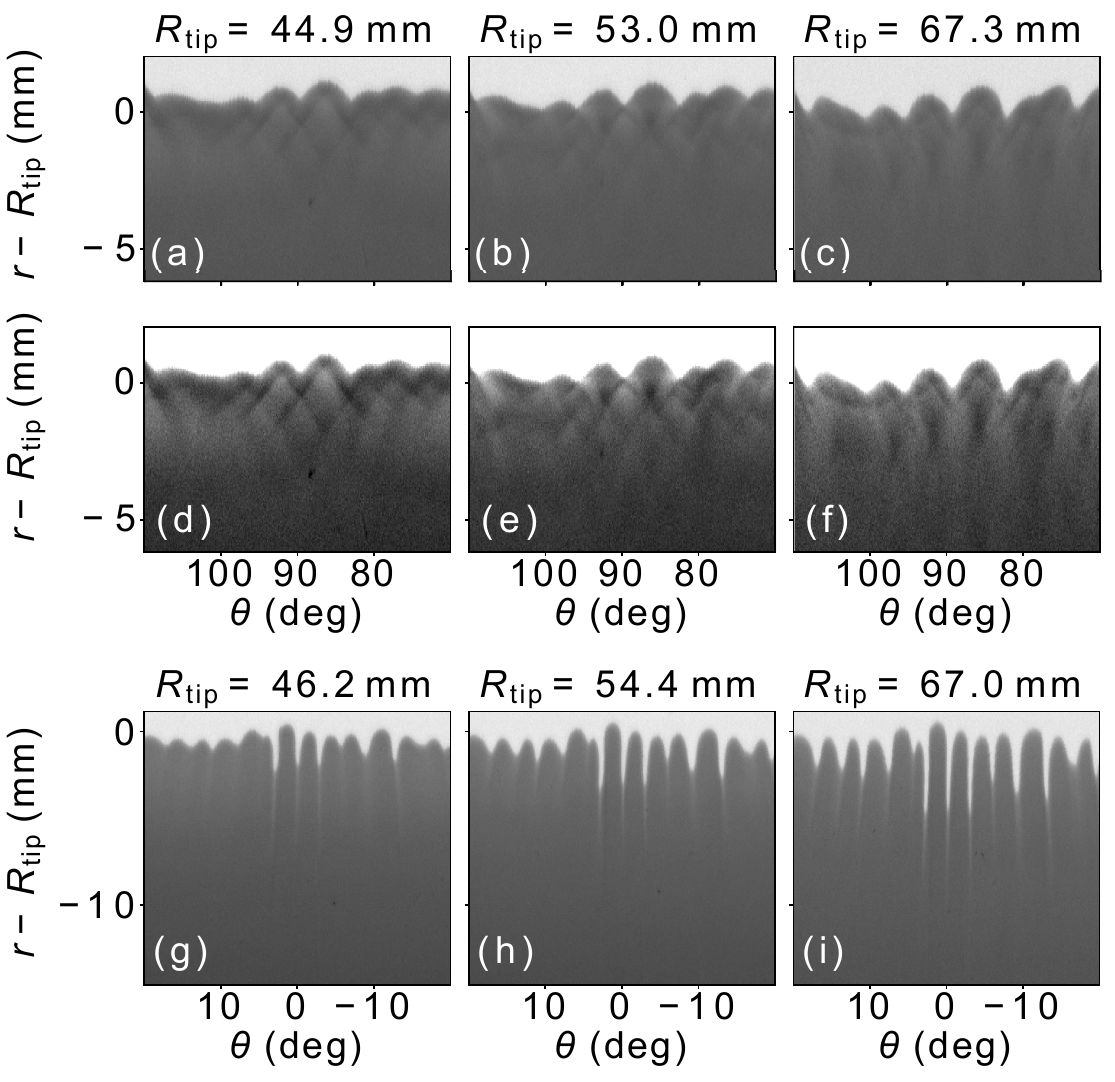}
    \caption{ Finger edges are tilted in the perpendicular growth direction but remain parallel to one another in the parallel direction. Images are mapped from the circular images onto Cartesian axes. (a-c) In the perpendicular growth direction, finger edges are tilted. (d-f) Triangular traces appear as darker lines that extend in back of the interface from the finger edges and cross one another. Same snapshots as (a–c) with enhanced contrast to show the traces more clearly. (g-i) For fingers growing parallel to the shear axis, the finger edges remain parallel, similar to the case without shear. (Fig.~\ref{fig:SI_noshear}).}
    \label{fig:shape}
\end{figure}

Shear not only widens fingers in the perpendicular direction, but also changes finger morphology as shown in Fig.~\ref{fig:shape} (which is mapped from circular to Cartesian axes). At $d_\mathrm{m}=6~\mathrm{mm}$ and $V_\mathrm{s}=4.3~\mathrm{mm/s}$, fingers in the perpendicular direction develop inclined sides (Fig.~\ref{fig:shape}a-c), while they have nearly parallel sides in the parallel direction (Fig.~\ref{fig:shape}g-i) or when there is no shear (Fig.~\ref{fig:SI_noshear}). Triangular-wave-like traces are observed behind the front as darker lines that extend from the finger edges and cross each other as seen in Fig.~\ref{fig:shape}d; these features gradually fade away as the front propagates forward and as the number of shear cycles increases (Fig.~\ref{fig:shape}f).

For fingers propagating perpendicularly to the shear axis, the transverse direction is along the shear axis. The stronger amplitude dependence of $W$ suggests that the shear smears different layers of the inner fluid in the $z$ direction by different amounts and the separated layers can overlap with each other to form darker traces. The concentration measurements shown below test whether the broader finger width is accompanied by a spread of inner fluid along the direction transverse to the finger growth. 

\subsection*{Smeared concentration profile}
\begin{figure}
    \centering
    \includegraphics[width=1\linewidth]{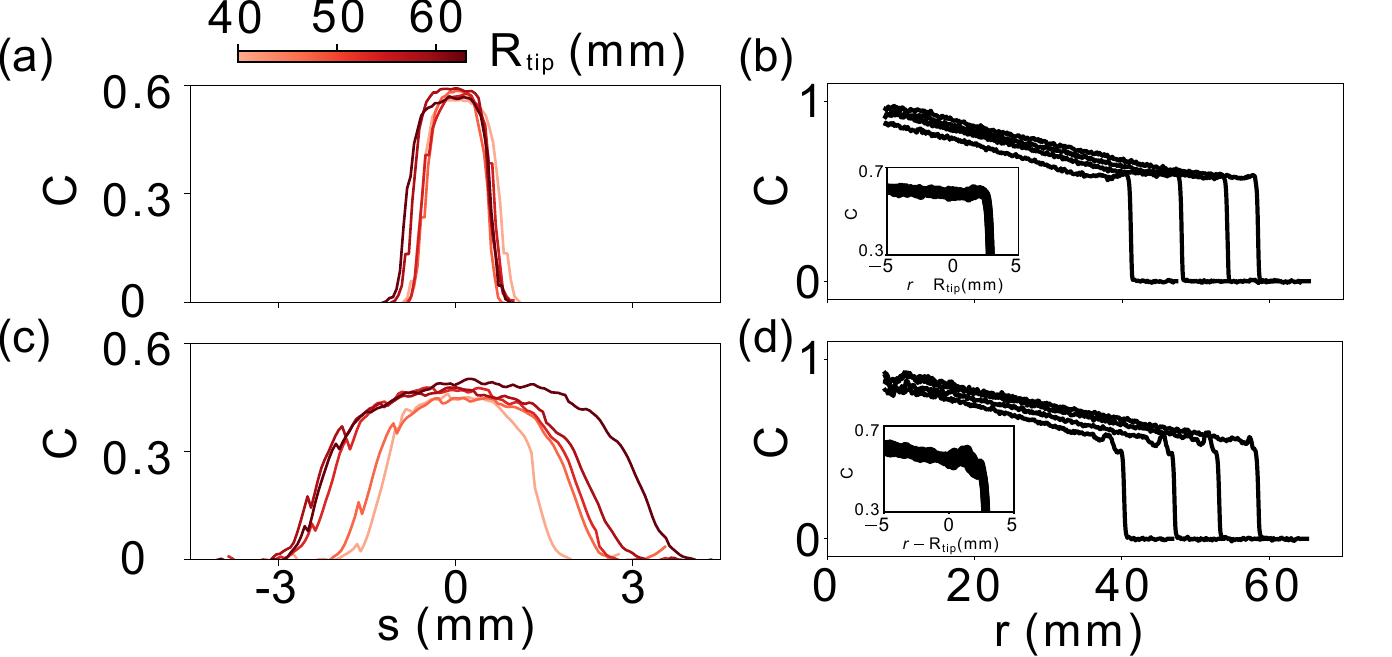}
    \caption{Concentration profiles in azimuthal and radial directions. (a,b) without shear and (c,d) with shear 
$d_\mathrm{m} = \mathrm{6~mm}$, $V_\mathrm{s} = \mathrm{4.3~mm/s}$. (a,c) Azimuthal concentration profiles 
with inner fluid concentration, $C$, versus arc length $s$. The cut is taken halfway along the finger, at
$r = R_\mathrm{tip} - R_\mathrm{f}/2$. The finger becomes wider with more gradual edges when there is shear. (b,d) Radial concentration profiles. With shear, the radial profiles develop an additional small peak just behind the tip as shown in the insets.}

    \label{fig:concentration}
\end{figure}

Figure~\ref{fig:concentration} compares representative perpendicular profiles without shear and with shear when $d_\mathrm{m}=6~\mathrm{mm}$ and $V_\mathrm{s}=4.3~\mathrm{mm/s}$, which are the same parameters as in Fig.~\ref{fig:shape}. The azimuthal cut is taken halfway along a finger, at $R_\mathrm{tip}-R_\mathrm{f}/2$. The arc length is measured relative to the finger center. 
Without shear, the profiles along the transverse direction are relatively narrow and have sharp sides (Fig.~\ref{fig:concentration}a).  These transitions become more gradual when there is shear as seen in Fig.~\ref{fig:concentration}c. This is also shown by the derivatives in Fig.~\ref{fig:SI_edge} where the derivative at the finger edges decreases with increasing  $d_\mathrm{m}$. The broader concentration distribution which occurs when there is shear is consistent with the increase of $W$ in the perpendicular direction in Fig.~\ref{fig:width}. 

The radial profiles reveal a second change. When there is shear, the profiles develop a peak immediately behind the tips as seen in Fig.~\ref{fig:concentration}d. This structure is qualitatively consistent with the triangular traces seen in Fig.~\ref{fig:shape}d-f.

The more gradual edge transition, the peak behind the tip in the radial concentration profile, and the triangular traces behind the interface suggest a differential displacement of the finger structure in the third dimension, $z$, across the gap. Fluid layers at different $z$ respond differently to the displacement of the top plate. The separate motion of different layers can broaden the projected azimuthal profile and make the transition at the edge more gradual.  At the same time, overlapping different layers can generate triangular traces and give rise to additional structure behind the tip. This interpretation emphasizes the importance of the gap structure previously established in miscible Hele-Shaw displacement~\cite{lajeunesse1999,oliveira2011,lister2025,luo2020}. The triangular traces also confirm that the smoother gap-averaged profile is not caused by molecular mixing, which was minimized by working at high $\mathrm{Pe>1750}$.

\subsection*{Reconnection of fingers}
Since shear mainly deforms perpendicular fingers in the azimuthal direction, this raises the question of how it suppresses finger growth in the radial direction. Under some shear conditions, neighboring fingers reconnect, and these cases provide a window for studying how azimuthal deformation alters the growth in the radial direction. 

\begin{figure}
    \centering
    \includegraphics[width=1.02\linewidth]{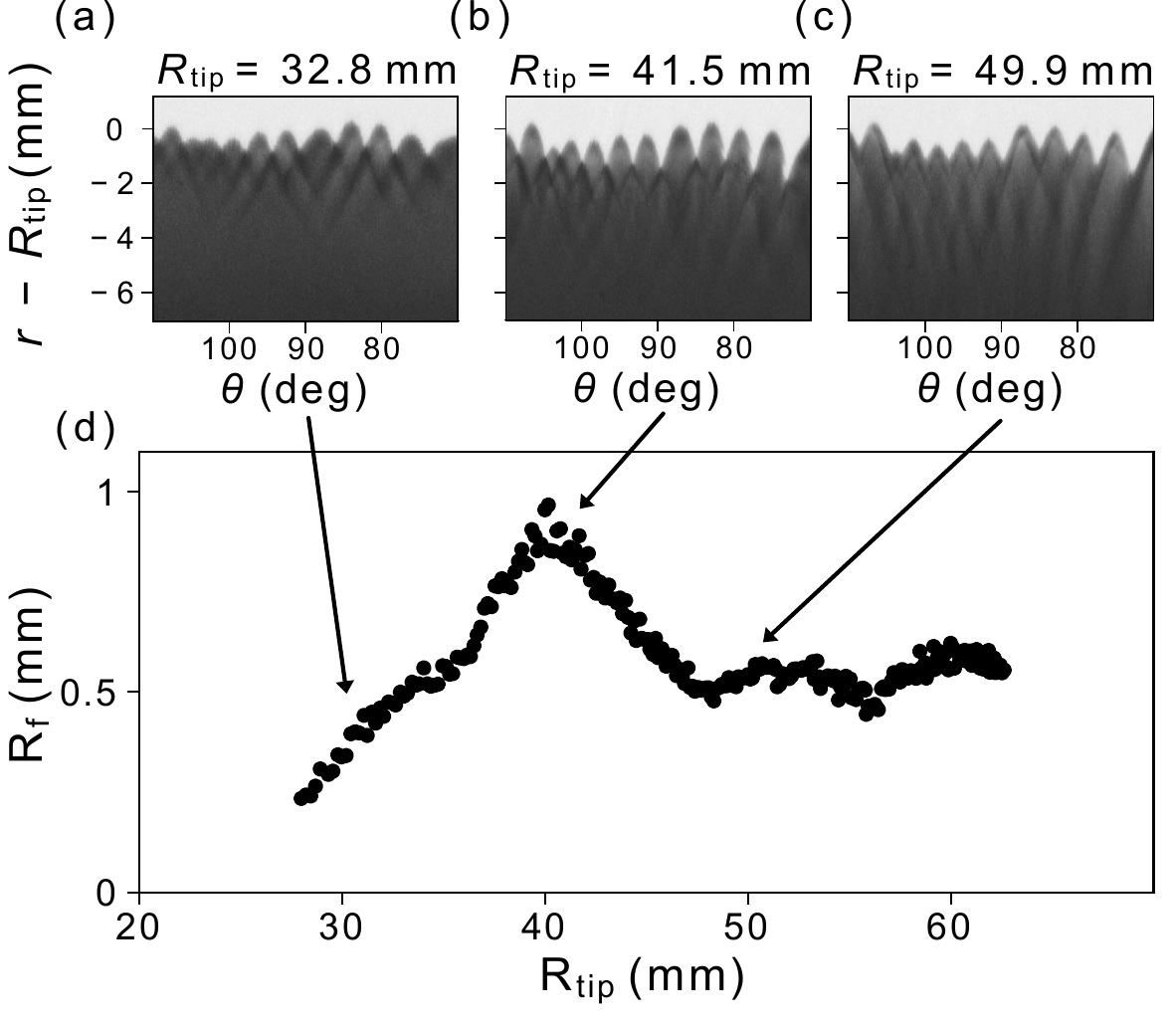}
    \caption{Example of finger reconnection. (a-c) $r - R_\mathrm{tip}$ is plotted versus angle, taken (a) before reconnection, (b) close to the point of reconnection and (c) after the reconnection occurs. Each image shows radius relative to the finger tip. Fingers grow longer from (a) to (b), but reconnect to be shorter from (b) to (c). (d) Finger length, $R_\mathrm{f}$, versus finger tip radius, $R_\mathrm{tip}$. $R_\mathrm{f}$ increases until $R_\mathrm{tip} \approx \mathrm{40}$~mm, and then drops as the fingers reconnect. Arrows connect each snapshot in (a)–(c) to the corresponding point in (d). $d_\mathrm{m}=\mathrm{6~mm}$, $V_\mathrm{s}=\mathrm{2.5~mm/s}$.}
    \label{fig:connect}
\end{figure}

Figure~\ref{fig:connect} shows a representative interface evolution pattern at $d_\mathrm{m}=6~\mathrm{mm}$ and $V_\mathrm{s}=2.5~\mathrm{mm/s}$, where a more detailed evolution is shown in Fig.~\ref{fig:SI_reconnect}.  In the images mapped into polar coordinates, the fingers initially grow as the interface advances but subsequently come into contact with their neighbors, reducing the radial distance between the tips and the valleys. The fingers therefore become shorter without any inward motion of the tips. The plot of $R_\mathrm{f}$ versus $R_\mathrm{tip}$ (Fig.~\ref{fig:connect}d) captures this shortening, and the flatter curve after reconnection shows that the fingers also grow more slowly afterward. In these cases, azimuthal widening affects radial growth in two ways: it shortens existing fingers through reconnection, and it reduces their subsequent growth rate.

\begin{figure}
    \centering
    \includegraphics[width=1\linewidth]{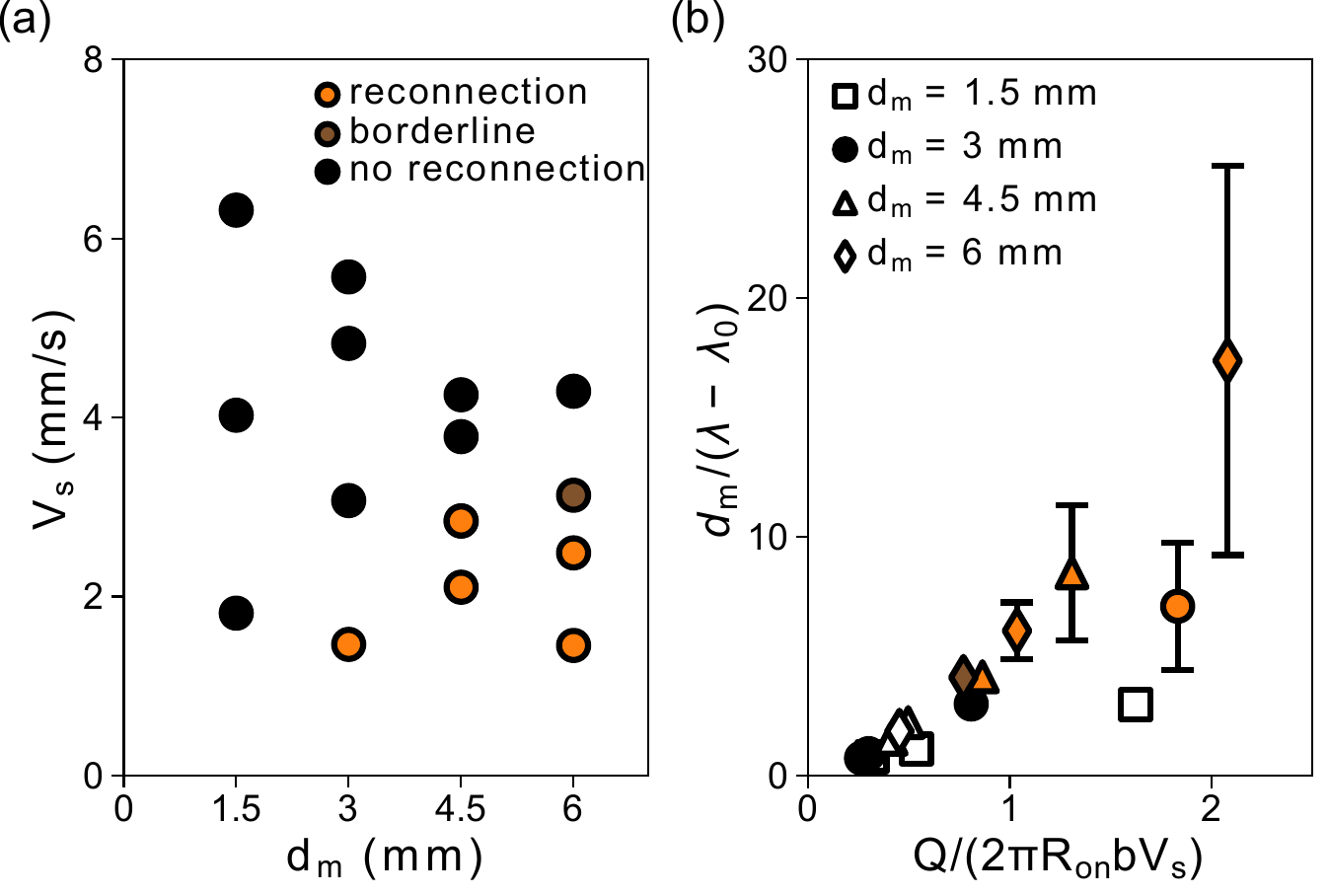}
    \caption{Conditions for fingers to reconnect. (a) Phase diagram of whether reconnection happens for different shear speed $V_\mathrm{s}$ and shear amplitude $d_\mathrm{m}$. Neither $V_\mathrm{s}$ nor $d_\mathrm{m}$ alone determines whether fingers reconnect. (b) Same experiments plotted as a competition between shear amplitude and finger wavelength at the onset, $d_\mathrm{m}/(\lambda-\lambda_0)$, versus a competition between finger propagation speed at the onset and shear speed, $Q/(2\pi R_\mathrm{on} b V_\mathrm{s})$. Reconnection happens as $d_\mathrm{m}/(\lambda-\lambda_0)\gtrsim 5$. The classification of experiments with reconnection or at the boundary is shown in Fig.~\ref{fig:SI_growth}.
}
    \label{fig:conditions}
\end{figure}

To understand why reconnection happens, we examine the conditions under which it is observed, as shown in Fig.~\ref{fig:conditions}a. Reconnection occurs at larger $d_\mathrm{m}$ and lower $V_\mathrm{s}$. Neither amplitude nor speed alone determines the outcome. As shear widens the finger laterally and part of this widening has already occurred at the onset by $\lambda-\lambda_\mathrm{0}$, where $\lambda=W\cdot R_\mathrm{on}$ is the linear finger wavelength and $\lambda_\mathrm{0}$ is the critical wavelength when there is no shear. For fingers to reach their neighbors, the shear amplitude must therefore be large enough compared with the widening that has already occurred by onset, so that the fingers can continue to get wider afterward and touch their neighbors.

Figure~\ref{fig:conditions}b compares these two length scales and shows that reconnection occurs only when the shear amplitude exceeds approximately five times the excess onset wavelength. Figure~\ref{fig:SI_growth} shows $R_\mathrm{f}$ versus $R_\mathrm{tip}$ for experiments that take place at or above the borderline where reconnection occurs. Also included for comparison is one example in which reconnection does not occur.

\section*{Discussion and Conclusion}
Oscillatory translational shear delays the fingering onset and reduces subsequent finger growth for fingers propagating perpendicular to the shear axis, similar to what was observed in the parallel direction~\cite{liu2026}. Despite this similar response, the morphology is significantly different. Fingers in the perpendicular directions widen laterally, whereas those in the parallel directions become narrower and more elongated. In both directions, shear deforms the structure in the $z$-direction within the gap in the direction along the shear axis. For perpendicular fingers this is in the direction transverse to the finger growth and for fingers growing in the direction parallel to the shear axis this is in the longitudinal direction of the finger growth. 

The inner-fluid concentration measurements provide further evidence of this lateral deformation. Perpendicular fingers develop broader transverse profiles with more gradual edges, suggesting smearing of the fluid across the lateral direction. The additional peak structure behind the tips also suggests a differential displacement of the finger structure at different layers in the $z$ direction across the gap. Together with the width measurements, these profiles indicate that shear changes the lateral extent of individual fingers and therefore enables fingers to interact more readily with their neighbors. 

When shear amplitude is greater than approximately five times the excess finger wavelength at the onset, this lateral deformation brings neighboring structures into contact and produces reconnection. Reconnection reduces the finger length when the finger tips subsequently continue to advance outward. Finger shortening therefore results from a change in the intervening valleys rather than tip retraction. After reconnection, finger length also grows more slowly. 

Although reconnection does not account for all of the suppression, as it only happens for a subset of shear parameters and after the onset, we can propose a similar process before the onset. Before fingers become visible, small-amplitude perturbations are developed along the interface. Similar to the reconnection of fingers, shear can also connect small-amplitude perturbations with their neighbors, and thus reduce the amplitude of the perturbations as well as suppress the development of the instability. This connection of neighboring perturbations can also serve as a possible explanation for the increase of finger width. Testing this hypothesis will require a quantitative stability analysis.

Together, our results show that shear influences both the appearance of fingers and their subsequent evolution. They support a physical picture in which transverse shear smears fingers laterally, modifies their interactions with neighbors, and thereby affects radial pattern development. Reconnection provides a particularly clear example of this coupling. Because it occurs after onset, it explains an observed route to reduced radial protrusion during later growth.

\section*{Acknowledgments}
I thank  Samar Alqatari, Irmgard Bischofberger, Justin C. Burton, Varda F. Hagh,  Savannah D. Gowen, Heinrich M. Jaeger, Chloe W. Lindeman, Neomie Livne,  Ayanna Matthews, Nidhi Pashine, Alice Pelosse, Robert Rosner, David W. Schmitz, Thomas E. Videb{\ae}k, and Thomas A. Witten. I would especially acknowledge Sidney R. Nagel for his contribution throughout the development of this work and my development as a scientist.

Claude was used for grammar correction. This work was supported by United States Department of Energy Basic Energy Sciences Grant DE-SC0020972. 

\section*{DATA AVAILABILITY}
The data that support the findings of this article are included in the main text and Supplementary Information.

\bibliography{apssamp}

@article{saffman1958,
 author={Saffman, P. G. and Taylor, G. I.},
 title={The penetration of a fluid into a porous medium or {Hele-Shaw} cell containing a more viscous liquid},
 journal={Proc. R. Soc. London, Ser. A}, volume={245}, pages={312--329}, year={1958},
 doi={10.1098/rspa.1958.0085}
}

@article{paterson1981,
 author={Paterson, L.}, title={Radial fingering in a {Hele Shaw} cell},
 journal={J. Fluid Mech.}, volume={113}, pages={513--529}, year={1981},
 doi={10.1017/S0022112081003613}
}

@article{lajeunesse1999,
 author={Lajeunesse, E. and Martin, J. and Rakotomalala, N. and Salin, D. and Yortsos, Y. C.},
 title={Miscible displacement in a {Hele-Shaw} cell at high rates},
 journal={J. Fluid Mech.}, volume={398}, pages={299--319}, year={1999},
 doi={10.1017/S0022112099006357}
}

@article{bischofberger2014,
 author={Bischofberger, I. and Ramachandran, R. and Nagel, S. R.},
 title={Fingering versus stability in the limit of zero interfacial tension},
 journal={Nat. Commun.}, volume={5}, pages={5265}, year={2014},
 doi={10.1038/ncomms6265}
}

@article{oliveira2011,
 author={Oliveira, R. M. and Meiburg, E.},
 title={Miscible displacements in {Hele-Shaw} cells: three-dimensional {Navier-Stokes} simulations},
 journal={J. Fluid Mech.}, volume={687}, pages={431--460}, year={2011},
 doi={10.1017/jfm.2011.367}
}

@article{lister2025,
 author={Lister, J. R. and Dauck, T.-F.},
 title={Fingering instability of self-similar radial flow of miscible fluids in a {Hele-Shaw} cell},
 journal={J. Fluid Mech.}, volume={1008}, pages={A31}, year={2025},
 doi={10.1017/jfm.2025.153}
}

@article{videbaek2019,
 author={Videb{\ae}k, T. E. and Nagel, S. R.},
 title={Diffusion-driven transition between two regimes of viscous fingering},
 journal={Phys. Rev. Fluids}, volume={4}, pages={033902}, year={2019},
 doi={10.1103/PhysRevFluids.4.033902}
}

@article{zheng2015,
 author={Zheng, Z. and Kim, H. and Stone, H. A.},
 title={Controlling viscous fingering using time-dependent strategies},
 journal={Phys. Rev. Lett.}, volume={115}, pages={174501}, year={2015},
 doi={10.1103/PhysRevLett.115.174501}
}

@article{gao2019,
 author={Gao, T. and Mirzadeh, M. and Bai, P. and Conforti, K. M. and Bazant, M. Z.},
 title={Active control of viscous fingering using electric fields},
 journal={Nat. Commun.}, volume={10}, pages={4002}, year={2019},
 doi={10.1038/s41467-019-11939-7}
}

@article{liu2026,
 author={Liu, Z. and Alqatari, S. and Videb{\ae}k, T. E. and Nagel, S. R.},
 title={Effect of translational shear on interfacial structure in the viscous fingering instability},
 journal={Sci. Adv.}, volume={12}, pages={eaeb2907}, year={2026},
 doi={10.1126/sciadv.aeb2907}
}

@article{Homsy1987,
  author = {Homsy, G. M.},
  title = {Viscous fingering in porous media},
  journal = {Annu. Rev. Fluid Mech.},
  volume = {19},
  pages = {271--311},
  year = {1987},
  doi = {10.1146/annurev.fl.19.010187.001415}
}

@article{TanHomsy1988,
  author = {Tan, C. T. and Homsy, G. M.},
  title = {Simulation of nonlinear viscous fingering in miscible displacement},
  journal = {Phys. Fluids},
  volume = {31},
  number = {6},
  pages = {1330--1338},
  year = {1988},
  doi = {10.1063/1.866726}
}

@article{MirandaWidom1998,
  author = {Miranda, Jos{\'e} A. and Widom, Michael},
  title = {Radial fingering in a {Hele-Shaw} cell: A weakly nonlinear analysis},
  journal = {Physica D},
  volume = {120},
  number = {3--4},
  pages = {315--328},
  year = {1998},
  doi = {10.1016/S0167-2789(98)00097-9}
}

@article{KesslerLevine1986,
  author = {Kessler, David A. and Levine, Herbert},
  title = {Coalescence of {Saffman-Taylor} fingers: A new global instability},
  journal = {Phys. Rev. A},
  volume = {33},
  number = {5},
  pages = {3625(R)--3627(R)},
  year = {1986},
  doi = {10.1103/PhysRevA.33.3625}
}

@article{Li2009,
  author = {Li, Shuwang and Lowengrub, John S. and Fontana, Jake and Palffy-Muhoray, Peter},
  title = {Control of viscous fingering patterns in a radial {Hele-Shaw} cell},
  journal = {Phys. Rev. Lett.},
  volume = {102},
  number = {17},
  pages = {174501},
  year = {2009},
  doi = {10.1103/PhysRevLett.102.174501}
}

@article{Dias2012,
  author = {Dias, Eduardo O. and Alvarez-Lacalle, Enrique and Carvalho, M{\'a}rcio S. and Miranda, Jos{\'e} A.},
  title = {Minimization of viscous fluid fingering: A variational scheme for optimal flow rates},
  journal = {Phys. Rev. Lett.},
  volume = {109},
  number = {14},
  pages = {144502},
  year = {2012},
  doi = {10.1103/PhysRevLett.109.144502}
}

@article{MirzadehBazant2017,
  author = {Mirzadeh, Mohammad and Bazant, Martin Z.},
  title = {Electrokinetic control of viscous fingering},
  journal = {Phys. Rev. Lett.},
  volume = {119},
  number = {17},
  pages = {174501},
  year = {2017},
  doi = {10.1103/PhysRevLett.119.174501}
}

@article{Chui2015.92.041003,
  title = {Interface evolution during radial miscible viscous fingering},
  author = {Chui, Jane Y. Y. and de Anna, Pietro and Juanes, Ruben},
  journal = {Phys. Rev. E},
  volume = {92},
  issue = {4},
  pages = {041003(R)},
  numpages = {5},
  year = {2015},
  month = {Oct},
  publisher = {American Physical Society},
  doi = {10.1103/PhysRevE.92.041003},
  url = {https://link.aps.org/doi/10.1103/PhysRevE.92.041003}
}

@article{luo2020,
  title = {Particle-induced miscible fingering: Continuum limit},
  author = {Luo, Rui and Chen, Yun and Lee, Sungyon},
  journal = {Phys. Rev. Fluids},
  volume = {5},
  issue = {9},
  pages = {094301},
  numpages = {18},
  year = {2020},
  month = {Sep},
  publisher = {American Physical Society},
  doi = {10.1103/PhysRevFluids.5.094301},
  url = {https://link.aps.org/doi/10.1103/PhysRevFluids.5.094301}
}

@article{Pihler2012,
  title = {Suppression of Complex Fingerlike Patterns at the Interface between Air and a Viscous Fluid by Elastic Membranes},
  author = {Pihler-Puzovi{\ifmmode \acute{c}\else \'{c}\fi{}}, D. and Illien, P. and Heil, M. and Juel, A.},
  journal = {Phys. Rev. Lett.},
  volume = {108},
  issue = {7},
  pages = {074502},
  numpages = {5},
  year = {2012},
  month = {Feb},
  publisher = {American Physical Society},
  doi = {10.1103/PhysRevLett.108.074502},
  url = {https://link.aps.org/doi/10.1103/PhysRevLett.108.074502}
}

@article{cross1993,
  title = {Pattern formation outside of equilibrium},
  author = {Cross, M. C. and Hohenberg, P. C.},
  journal = {Rev. Mod. Phys.},
  volume = {65},
  issue = {3},
  pages = {851--1112},
  numpages = {0},
  year = {1993},
  month = {Jul},
  publisher = {American Physical Society},
  doi = {10.1103/RevModPhys.65.851},
  url = {https://link.aps.org/doi/10.1103/RevModPhys.65.851}
}

@article{
Qing2023,
author = {Qing Zhang  and Shuang Zhou  and Rui Zhang  and Irmgard Bischofberger },
title = {Dendritic patterns from shear-enhanced anisotropy in nematic liquid crystals},
journal = {Science Advances},
volume = {9},
number = {2},
pages = {eabq6820},
year = {2023},
doi = {10.1126/sciadv.abq6820},
URL = {https://www.science.org/doi/abs/10.1126/sciadv.abq6820},
}

@article{
Langer1989,
author = {J. S. Langer },
title = {Dendrites, Viscous Fingers, and the Theory of Pattern Formation},
journal = {Science},
volume = {243},
number = {4895},
pages = {1150-1156},
year = {1989},
doi = {10.1126/science.243.4895.1150},
URL = {https://www.science.org/doi/abs/10.1126/science.243.4895.1150}}

\newpage
\clearpage

\renewcommand{\thefigure}{S\arabic{figure}}
\renewcommand{\thetable}{S\arabic{table}}
\renewcommand{\theequation}{S\arabic{equation}}
\renewcommand{\thepage}{S\arabic{page}}
\setcounter{figure}{0}
\setcounter{table}{0}
\setcounter{equation}{0}
\setcounter{page}{1}

\section*{Supplementary Information}
\subsection*{Methods}

The apparatus and experimental protocols follow Ref.~\cite{liu2026} and are summarized here. The Hele-Shaw cell is constructed by two circular glass plates with radii $R=14~\mathrm{cm}$ and a thickness greater than $1.27~\mathrm{cm}$ to prevent bending. The gap spacing was kept uniform at $b=305~\mathrm{\mu m}$ by six spacers equally placed around the periphery of the plates. To ensure a uniform gap during shear, the top plate is maintained closely in contact with the spacers by weights equally placed around the perimeter. The bottom plate rests on three T-slotted framing rails, where screws hold it in place and bumpers damp vibration. An aluminum ring was used to enclose the top plate and connect it to a 2-inch Feedback Rod Linear Actuator (FA-PO-35-12-2 from Firgelli Automations). The actuator was driven back and forth with constant speed and amplitude by a programmed Arduino with a High Power Motor Drive (HiLetgo BTS7960 43A). There is an inlet at the center of the top plate with a diameter of $1.6~\mathrm{mm}$, to which both the injection tube and the waste tube were connected via a Y-shaped tubing connector. The bottom plate is flat without any holes, so no perturbation to the flow can be caused by an opposing hole during the applied shear.

Before each run, the two plates were carefully aligned, and the outer fluid was injected to the edge of the plates. Before injecting the inner fluid, bubbles and the outer fluid residual were removed by flushing the inner fluid through the injection tube to the inlet and then pumping it out from the waste tube. The valve for the waste tube was then closed, and then the inner fluid was injected at the same time as shear was turned on, using a syringe pump (NE-1000 from New Era Pump Systems Inc.).

Both the inner and outer fluids are mixtures of glycerol and water whose viscosities are $\eta_{\rm in}= 35 \pm 2~\rm mPa\cdot s$ and $\eta_{\rm out}= 218 \pm 5~\rm mPa\cdot s$ respectively. Their viscosities were measured by an Anton Paar MCR 301 rheometer. The inner fluid was dyed with $0.04\%$ (wt) of brilliant blue G-250 ( Alfa Aesar), so that we can measure the concentration, $C$, from calibrated intensity of transmitted light ~\cite{liu2026,videbaek2019,bischofberger2014}. The radial concentration profile $C(r)$ is taken along the line from the center of the pattern to the tip of the interface. The azimuthal concentration profile $C(s)$ is tracked along the parallel direction to the shear axis, and measured at a distance that is halfway between the finger tip and the radius of the valley between fingers: $R_\mathrm{tip}-1/2R_\mathrm{f}$.

The top plate displacement was measured relative to the stationary bottom plate using two parallel lines perpendicular to the shear axis, one on each plate. As the maximum displacement was not captured by the camera with a finite frame rate, $d_\mathrm{m}$ was measured as half the difference between two maximum excursions centered around the beginning point over multiple cycles with an error of $0.06~\mathrm{mm}$. We calculate the shear speed $V_\mathrm{s}$ as $4d_\mathrm{m}/T_\mathrm{s}$, where $T_\mathrm{s}$ is the period averaged over multiple cycles. So the error of $V_\mathrm{s}$ is less than $3.6\%$.

A Prosilica GX 3300 camera below the bottom plate records both the fingering pattern and the plate motion. We unwrapped the interface between the two fluids into polar coordinates with $\theta = 0$ ($\theta = \pi$) along the shear axis. To determine how the onset radius varies with direction, we divide the interface into 32 equal angular segments, each spanning $\pi/16$, with the first segment centered at $\theta = 0$. The onset radius $R_{\mathrm {on}}$ is measured separately in each segment. Within each segment, onset is defined by the first appearance of a local minimum whose depth relative to the nearest local maximum is at least $0.2~\mathrm{mm}$. The radial position of this minimum on the original interface gives $R_{\mathrm {on}}$ for that segment. To reduce the influence of noise, we fit $R_{\rm on}(\theta)$ with a Fourier series truncated at second order, which captures the twofold symmetry imposed by uniaxial shear. The fitted values are then used as the onset radii in the parallel and perpendicular directions.

The angular finger width $W$ is obtained in a similar way. We measured the angular width of every individual finger as the angular separation between two minima on either side of a finger. The separation is measured around the full interface, $0 \le \theta < 2\pi$, as $W(\theta)$. We then fit $W(\theta)$ with a Fourier series truncated at second order, and the fitted values in the parallel and perpendicular directions are used as the finger widths in those directions.

We track individual fingers near the perpendicular directions and define their length $R_{\rm f}$ as the radial distance between a finger tip (a local maximum of the interface) and the average radius of the valleys on either side of it.

\subsection*{Bluntness at the tip of a finger in the radial direction does not change significantly under shear}
\begin{figure}
    \centering
    \includegraphics[width=1\linewidth]{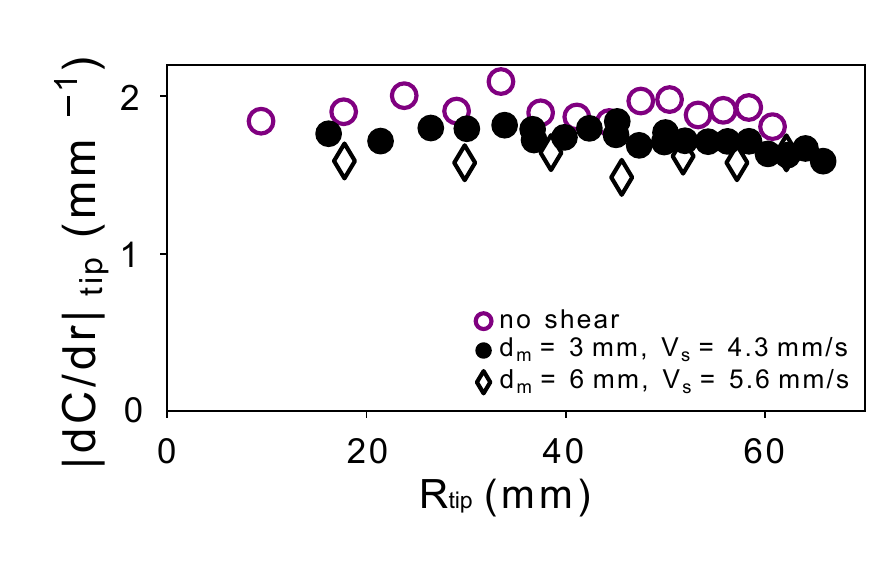}
    \caption{Derivative of radial concentration at the finger tip for fingers propagating perpendicular to the shear axis.  $\mathrm{d}C/\mathrm{d}r$, plotted versus the interface radius, $R_\mathrm{tip}$, as the interface expands. Larger values indicate a sharper radial transition and therefore a blunter interface across the gap. Data are shown without shear (open purple circles), with $d_\mathrm{m} = 3$~mm and $V_\mathrm{s} = 4.3$~mm/s (filled circles), and with $d_\mathrm{m} = 6$~mm and $V_\mathrm{s} = 5.6$~mm/s (open diamonds). $Q = 133~\mu\mathrm{L/s}$ in all experiments.}

    \label{fig:SI_tip}
\end{figure}

To test whether shear smooths the interface radially for fingers propagating perpendicular to the shear axis, we measured the derivative of radial concentration profile right at the finger tip, $|\mathrm{d}C/\mathrm{d}r|_\mathrm{tip}$, against the interface radius, $R_\mathrm{tip}$. A larger $|\mathrm{d}C/\mathrm{d}r|_\mathrm{tip}$ corresponds to a sharper transition and thus a blunter interface spanning the gap~\cite{liu2026}. As shown in Fig.~\ref{fig:SI_tip}, $|\mathrm{d}C/\mathrm{d}r|_\mathrm{tip}$ remains nearly constant as the interface expands. Shear reduces it only slightly, from approximately $1.9~\mathrm{mm^{-1}}$ without shear to approximately $1.6~\mathrm{mm^{-1}}$ at $d_\mathrm{m} = 6$~mm. This change is small compared to that in the parallel direction, which decreases by roughly a factor of two over the same range of $d_\mathrm{m}$~\cite{liu2026}. Shear has little effect on the radial bluntness of the interface at the tip for fingers in the perpendicular direction.

\subsection*{Finger reconnection plot for different shear speeds and amplitudes}
\begin{figure}
    \centering
    \includegraphics[width=0.8\linewidth]{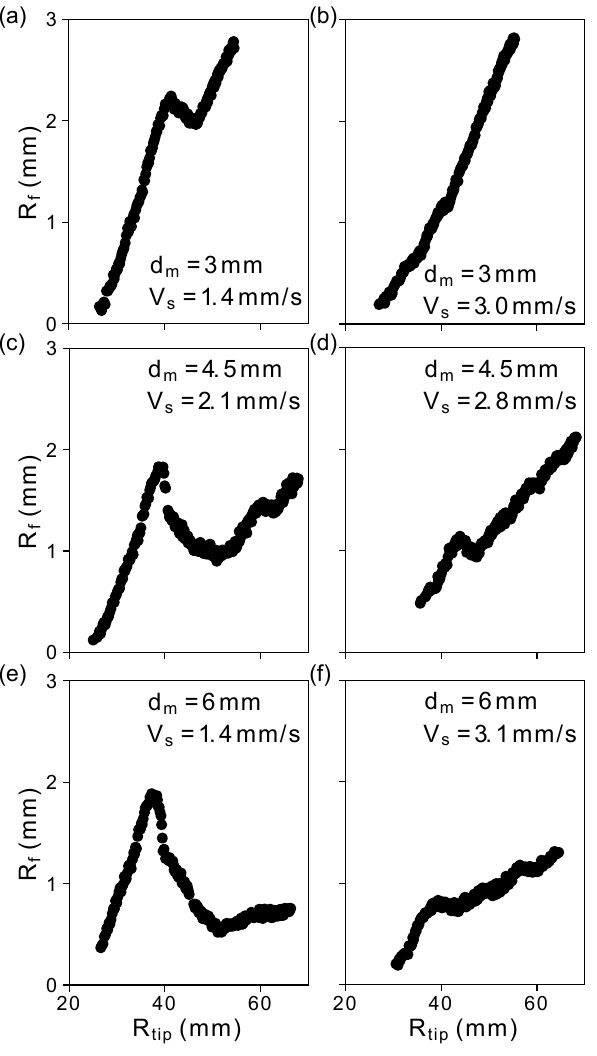}
    \caption{Finger growth in the perpendicular direction at different shear parameters to compare experiments with or without reconnection. (a,c,d,e) Fingers reconnect with a drop in finger length. (b) No reconnection. (f) Borderline case with unclear finger length drop and slower finger growth rate after around $R_\mathrm{tip}=40$ mm.}
    \label{fig:SI_growth}
\end{figure}

To show how reconnection depends on shear parameters, Fig.~\ref{fig:SI_growth} plots the finger length $R_\mathrm{f}$ against the tip radius $R_\mathrm{tip}$ for three shear amplitudes, $d_\mathrm{m} = 3$, $4.5$, and $6$~mm. When neighboring fingers reconnect, $R_\mathrm{f}$ drops; when they do not, $R_\mathrm{f}$ increases nearly linearly with $R_\mathrm{tip}$ as shown in Fig.~\ref{fig:SI_growth}b. One borderline example is shown in Fig.~\ref{fig:SI_growth}f, where a change of slope is observable but the peak is not prominent.

\subsection*{Effect of reconnection on finger shape compared with the no shear case}

\begin{figure*}
    \centering
    \includegraphics[width=1\linewidth]{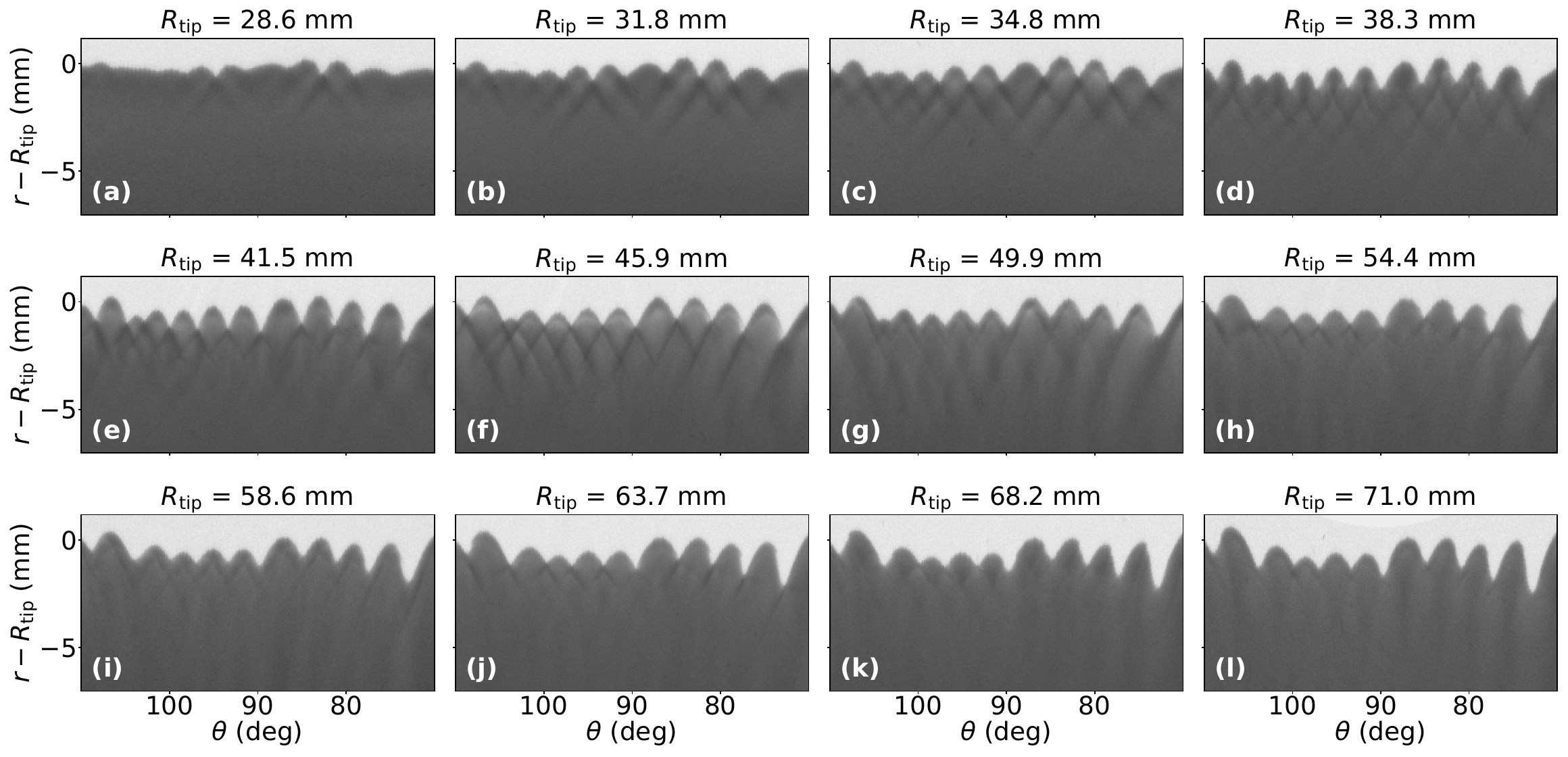}
    \caption{Evolution of the interface in the perpendicular direction in which neighboring fingers reconnect. $d_\mathrm{m} = 6$~mm, $V_\mathrm{s} = 2.5$~mm/s, $Q = 133~\mu\mathrm{L/s}$. Each panel shows $r - R_\mathrm{tip}$ plotted versus angle, $\theta$, in the perpendicular direction centered on $\theta = 90^\circ$. (a-l) are successive snapshots from $R_\mathrm{tip} = 28.6$ to $71.0$~mm. As the interface advances, the fingers take on a triangular shape and neighboring fingers come into contact, but become rounded at later frames.}
    \label{fig:SI_reconnect}
\end{figure*}

\begin{figure*}
    \centering
    \includegraphics[width=1\linewidth]{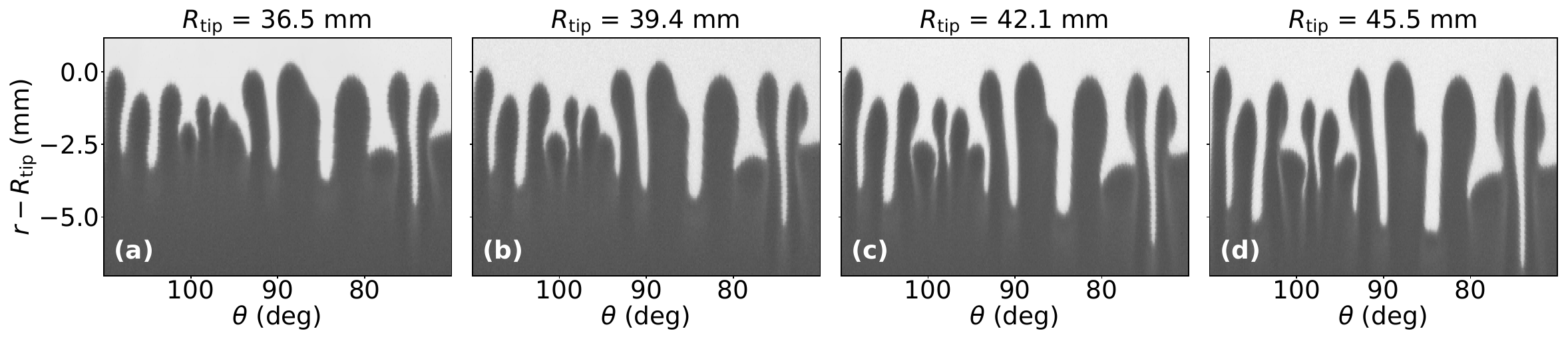}
    \caption{Evolution of the interface in the perpendicular direction without shear. (a-d) are snapshots from $R_\mathrm{tip} = 36.5$ to $45.5$~mm. The fingers have rounded tips and nearly parallel sides, in contrast to the triangular fingers when shear is present.}
    \label{fig:SI_noshear}
\end{figure*}

Figures~\ref{fig:SI_reconnect} and \ref{fig:SI_noshear} compare the evolution of the interface in the perpendicular direction with reconnection and without shear, shown on the same scale. When reconnection occurs (Fig.~\ref{fig:SI_reconnect}), fingers develop a triangular shape. At later frames, the traces fade and the finger tips become rounded again. In comparison, the fingers have rounded tips and nearly parallel sides when there is no shear (Fig.~\ref{fig:SI_noshear}).

\subsection*{Steepness at the edges of a finger at different shear speeds and amplitudes}
\begin{figure*}
    \centering
    \includegraphics[width=0.8\linewidth]{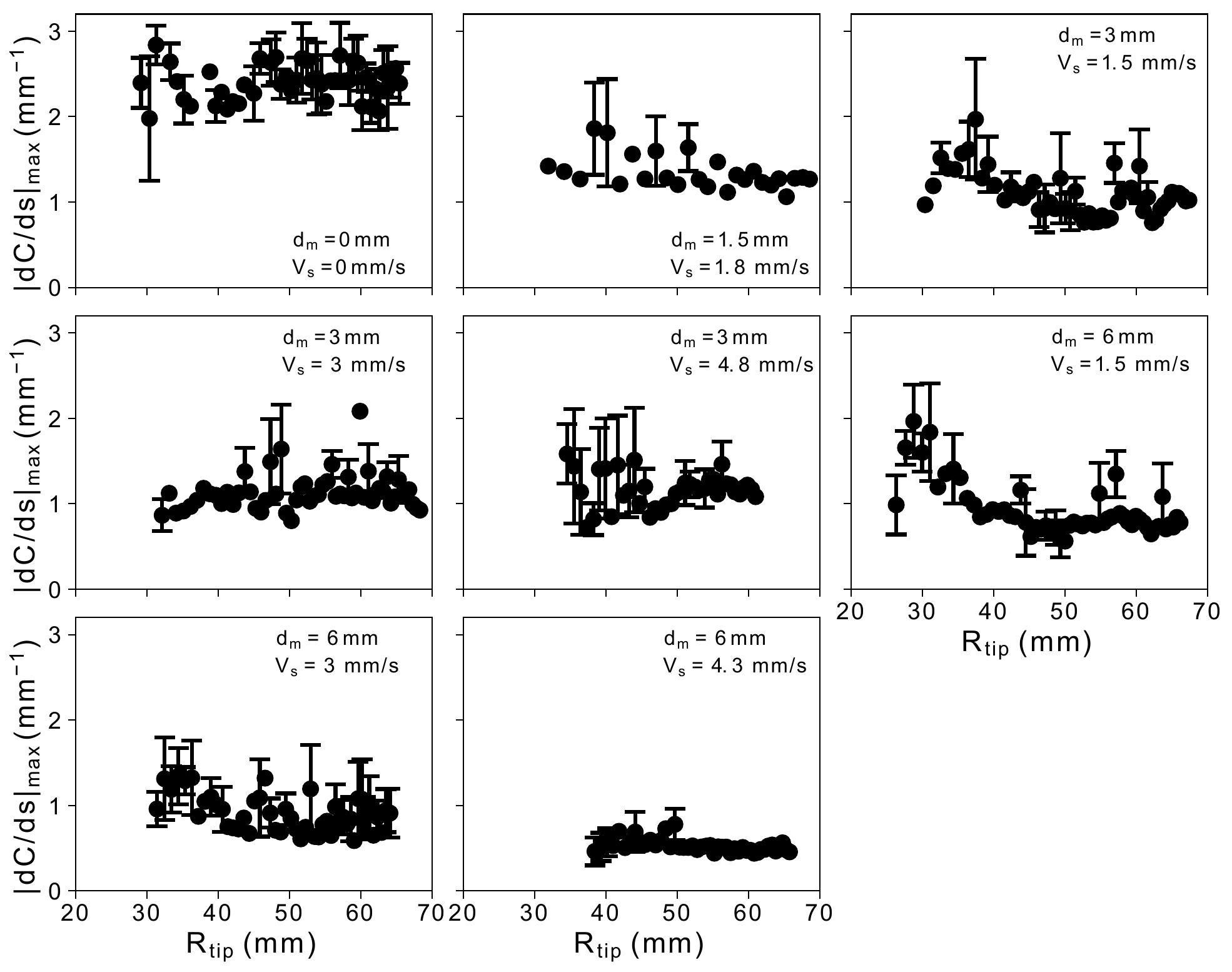}
    \caption{Steepest azimuthal concentration change at the edges of a finger in the perpendicular direction as the interface expands. For each image, the concentration profile, $C$, across one finger is taken along the azimuthal arc length $s$ as in Fig.~\ref{fig:concentration}a,c. $\mathrm{d}C/\mathrm{d}s$ is the steepest derivative at the finger edges. Smaller values correspond to more gradual edges. Each panel is one experiment, with the shear amplitude $d_\mathrm{m}$ and shear speed $V_\mathrm{s}$ indicated. $Q = 133~\mu\mathrm{L/s}$ in all experiments.}
    \label{fig:SI_edge}
\end{figure*}

To quantify how smeared the sides of the fingers are, we measured the steepest azimuthal concentration gradient, $\mathrm{d}C/\mathrm{d}s$, across individual fingers in the perpendicular direction (Fig.~\ref{fig:SI_edge}). Without shear, the edges are steep, with $|\mathrm{d}C/\mathrm{d}s|_\mathrm{max} \approx 2$--$3~\mathrm{mm^{-1}}$ as the interface at $R_\mathrm{tip}$ expands out. 
When shear is present, the derivative decreases as $d_\mathrm{m}$ increases: at large $R_\mathrm{tip}$ ($> 50$~mm), it is $1.2$--$1.5~\mathrm{mm^{-1}}$ for $d_\mathrm{m} = 1.5$~mm, $1~\mathrm{mm^{-1}}$ for $d_\mathrm{m} = 3$~mm, and $0.5$--$0.9~\mathrm{mm^{-1}}$ for $d_\mathrm{m} = 6$~mm. The effect of the shear speed is less clear. These data are consistent with the more gradual edges shown in Fig.~\ref{fig:concentration}.

\end{document}